\documentclass{article}

\begin{document}
\title{\bf World-line deviation and spinning particles}

\author{ Morteza Mohseni$^{\mbox{\small a,b,}}$\thanks{email: m-mohseni@pnu.ac.ir}\\
\\ \small $^{\mbox{\small a}}$Physics Department, Payame Noor
University, Tehran 19395-4697, Iran\\ \small $^{\mbox{\small b}}$
Institute for Studies in Theoretical Physics and Mathematics (IPM)
Tehran Iran.}
\maketitle

\begin{abstract}
A set of world-line deviation equations is derived in the
framework of  Mathisson-Papapetrou-Dixon desciption of
pseudo-classical spinning particles . They generalize the geodesic
deviation equations. We examine the resulting equations for
particles moving in the space-time of a plane gravitational wave.

\vspace{5mm}PACS: 04.20.-q; 04.25.-g;4.30.-Nk

\vspace{5mm}Keywords: geodesic deviation, spinning particles,
gravitational waves
\end{abstract}
\section{Introduction}
The geodesic deviation equation ,
\begin{eqnarray}\label{e1}
\frac{D^2 n^\mu}{D\tau^2}=-{R^\mu}_{\alpha\nu\beta}v^\alpha n^\nu
v^\beta
\end{eqnarray}
is one of the well-known equations of general relativity. This
equation shows the role of the space-time curvature on the motion
of test particles and has important applications , namely, it is
used for calculating relative accelerations of nearby particles in
an observer-independent manner, and may be integrated to give the
Lyapunov exponent in the study of chaotic behaviour of particle's
orbits.

There are several ways to derive this equation. One standard
derivation is as follows: Consider a two parameter family of
geodesic $x^\mu(\tau;\lambda)$ in which $\tau$ is the parameter
along the orbits and $\lambda$ characterize different orbits. To
obtain the separation of two nearby geodesics, one of them is
taken as "fiducial" geodesic described by, say, $x^\mu$, and the
other by  $x^\mu+\lambda n^\mu$ where $\lambda n^\mu$ represents
the separation. The equation for $n^\mu$ is then obtained by
inserting $x^\mu+\lambda n^\mu$ into the geodesic equation,
comparing terms linear in $\lambda$ and neglecting $O(\lambda^2)$.
This idea have been used in \cite{coli} to obtain generalized
geodesic deviation equations by considering expansions containing
higher orders of $\lambda$. These generalized equations have been
applied, for example, in \cite{coli} to the problem of closed
orbital motion of test particles in the Kerr space-time and in
\cite{van1} to the orbital motion in Schwarzchild metric. Other
interesting generalizations may be found in \cite{ciu} and
\cite{apr}. Another way to derive the equation is by varying a
suitable action \cite{baz}. This method was used in \cite{robs} to
generalize the equation to the so called "string deviation
equation". The quantization of the geodesic equation was discussed
in \cite{robq} and generalizing the equation to Kaluza-Klein
theories was done in \cite{mart}. Some other aspects or
applications of the equation have been discussed in \cite{mis}.

When forces other than gravity are present, or the particle has
some internal structure, it would no longer move along geodesics
in general. In these situations a "world-line deviation" equation
may be obtained by modifying the geodesic deviation equation by
taking the effect of the matter field or internal structure into
account. Thus using methods described in the previous paragraph, a
world-line deviation equation was obtained in \cite{bal} for the
of motion of charged particles in the framework of
Einstein-Maxwell theory. Another generalization of this type was
made in \cite{van2} (see also \cite{nieto}) for describing
spinning particles and was used there to study epicycles.

The equations used in \cite{van2} to determine the particle's
trajectories are a simplified version of
Mathisson-Papapertou-Dixon (MPD) equations \cite{dix}. The later
equations have been widely used to study the motion of spinning
particles (e.g. see \cite{moh} and references therein). The aim of
the present work is to obtain a world-line deviation in the
framework of MPD description of spinning particles. In the
following sections we apply the prescription mentioned earlier to
obtain the equations, and provide an example in which they could
be integrated analytically. Throughout the work, $1,2,3,4$ would
stand for $u,v,x,y$ as indices ,
${R^\mu}_{\nu\alpha\beta}=\partial_\alpha{\Gamma^\mu}_{\nu\beta}-
\partial_\beta{\Gamma^\mu}_{\nu\alpha}+{\Gamma^\mu}_{\alpha\delta}
{\Gamma^\delta}_{\nu\beta}-{\Gamma^\mu}_{\beta\delta}{\Gamma^\delta}_{\nu\alpha}$
, $\frac{D}{D\tau}$ and $\nabla$ represent covariant derivative,
the space-time signature would be $(-+++)$, and $[\mu,\nu]$ stands
for $\mu\nu-\nu\mu.$
\section{The equations}
The motion of a spinning particle is described by MPD equations
\begin{eqnarray}
\frac{Dp^\mu}{D\tau}&=&-\frac{1}{2}{R^\mu}_{\nu\kappa\lambda}v^\nu
s^{\kappa\lambda},\label{e2}\\
\frac{Ds^{\mu\nu}}{D\tau}&=&p^{[\mu} v^{\nu]},\label{e32}\\
p_\mu s^{\mu\nu}&=&0\label{e33}
\end{eqnarray}
in which $v^\mu=\frac{dx^\mu}{d\tau}$ is the particle's
four-velocity, $p^\mu$ its four-momentum and $s^{\mu\nu}$ its spin
tensor. We confine ourselves to time-like orbits with
\begin{eqnarray}\label{e3}
v_\mu v^\mu=-1.
\end{eqnarray}
MPD equations guarantee that the particle's mass and spin are
conserved
\begin{eqnarray}
p_\mu p^\mu&=&\mbox{constant}=-m^2,\label{e34}\\
\frac{1}{2}s_{\mu\nu}s^{\mu\nu}&=&\mbox{constant}=s^2.\label{e35}
\end{eqnarray}
Now consider a set $x^\mu(\tau;\lambda)$ describing the
world-lines of spinning particles of the same spin-to-mass ratios
and define
$$\frac{Ds^{\mu\nu}}{D\lambda}=J^{\mu\nu}$$ and
$$\frac{Dp^\mu}{D\lambda}=j^\mu.$$ Inserting $x^\mu(\tau)+\lambda n^\mu(\tau)$ into
(\ref{e2})-(\ref{e33}) and looking for $\lambda$ terms, we obtain
the following equations
\begin{eqnarray}\label{e5}
\frac{Dj^\mu}{D\tau}=-{R^\nu}_{\beta\alpha\kappa}v^\kappa n^\alpha
p^\beta-\frac{1}{2}{R^\mu}_{\nu\alpha\beta}\frac{Dn^\nu}{D\tau}s^{\alpha\beta}
-\frac{1}{2}{R^\mu}_{\nu\alpha\beta}v^\nu J^{\alpha\beta}
-\frac{1}{2}\nabla_\kappa {R^\mu}_{\nu\alpha\beta}n^\kappa v^\nu
s^{\alpha\beta}
\end{eqnarray}
\begin{eqnarray}\label{e4}
\frac{DJ^{\mu\nu}}{D\tau}=s^{\kappa[\mu}{R^{\nu]}}_{\kappa\alpha\beta}n^\alpha
v^\beta+p^{[\mu}\frac{Dn^{\nu]}}{D\tau}+j^{[\mu}v^{\nu]}
\end{eqnarray}
\begin{eqnarray}\label{e6}
s_{\mu\nu}j^\nu+J_{\mu\nu}p^\nu=0
\end{eqnarray}
respectively. If we turn off the spin, all terms except the first
in the right hand side of (\ref{e5}) vanish and the geodesic
deviation equation results. If we set $p^\mu=mv^\mu$ equations
(\ref{e5})-(\ref{e6}) reduce to those of \cite{van2}. It can be
seen from the above equations that there is no evolution equation
for $n^\mu$, it should be obtained indirectly from equations
(\ref{e5})-(\ref{e6}). The situation resembles the case of MPD
equations in which no direct equation exists for $v^\mu$. The
following equations
\begin{eqnarray}\label{e36}
v_\mu\frac{Dn^\mu(\tau)}{D\tau}=0,
\end{eqnarray}
\begin{eqnarray}\label{e8}
p_\mu j^\mu=0,
\end{eqnarray}
\begin{eqnarray} \label{e7}
s_{\mu\nu}J^{\mu\nu}=0.
\end{eqnarray}
are helpful in this regard. They stem from (\ref{e3})-(\ref{e35})
respectively. The latter two equations can be used for world-lines
describing nearby particles of the same spins and masses.
\section{The motion in a GW space-time}
In this section we consider the world-line deviations of spinning
particles in the space-time described by the following metric
\begin{eqnarray} \label{e9}
ds^2=-dudv-K(u,x,y) du^2+dx^2+dy^2
\end{eqnarray}
in which $u=t-z,v=t+z$ are light-cone coordinates and
$K(u,x,y)=f(u)(x^2-y^2)+2g(u)xy$. This metric represents a plane
gravitational wave of arbitrary polarization and profile
characterized by $f(u), g(u)$ propagating in $z$-direction. In
this space-time, the MPD equations admit the following solution
\begin{eqnarray}
v^\mu &=&(1,1,0,0),\nonumber\\
p^\mu &=&(m,m,0,0),\label{e11}\\
s^{1\mu}&=&s^{2\mu}=0,\hspace{3mm} s^{34}=\sigma\nonumber.
\end{eqnarray}
This solution describes a particle of mass $m$ and of spin
$\sigma$ ($s^\mu
:=\frac{1}{2\sqrt{-g}}{\epsilon^\mu}_{\nu\kappa\rho}p^\nu
s^{\kappa\rho}$, with $\epsilon^{1234}=-1$) sitting in the origin
of the coordinates with its spin along the $z$ direction. For the
above world-line, the geodesic deviation equation leads to the
following equations
\begin{eqnarray}
\frac{d^2n^3(\tau)}{d\tau^2}&=-f(u)n^3(\tau)-g(u)n^4(\tau),\label{e48}\\
\frac{d^2n^4(\tau)}{d\tau^2}&=f(u)n^4(\tau)-g(u)n^3(\tau).\label{e49}
\end{eqnarray}
We now consider two particles of the same spins and masses , one
initially at the origin and the other at a nearby point with a
specific separation. We aim to calculate this separation at any
value of the parameter $\tau$. We take (\ref{e11}) with
$(\tau,\tau,0,0)$ as the fiducial world-line. Now equation
(\ref{e6}) results in
\begin{eqnarray}
J^{12}(\tau)&=&0,\label{e13}\\J^{13}(\tau)+J^{23}(\tau)&=&-\frac{2\sigma}{m}j^4(\tau),
\label{e68}\\
J^{14}(\tau)+J^{24}(\tau)&=&\frac{2\sigma}{m}j^3(\tau)\label{e69}.
\end{eqnarray}
Equation (\ref{e7})-(\ref{e9}) results in
\begin{eqnarray}
\frac{dn^1(\tau)}{d\tau}+\frac{dn^2(\tau)}{d\tau}&=&0,\label{e41}\\
j^1(\tau)+j^2(\tau)&=&0,\label{e27}\\
J^{34}(\tau)&=&0\label{e14},
\end{eqnarray}
respectively. From equation (\ref{e5}) we obtain
\begin{eqnarray}
j^1(\tau)&=&\mbox{constant}=\alpha,\label{e16}\\
j^2(\tau)&=&\mbox{constant}=\beta,\label{e17}\\
\frac{dj^3(\tau)}{d\tau}&=&f(u)\left(J^{13}(\tau)-mn^3(\tau)\right)
+g(u)\left(J^{14}(\tau)-mn^4(\tau)\right),\label{e18}\\
\frac{dj^4(\tau)}{d\tau}&=&-f(u)\left(J^{14}(\tau)-mn^4(\tau)\right)
+g(u)\left(J^{13}(\tau)-mn^3(\tau)\right)\label{e19}.
\end{eqnarray}
Equation (\ref{e4}) leads to
\begin{eqnarray}
\frac{dJ^{12}(\tau)}{d\tau}&=&m\left(\frac{dn^2(\tau)}{d\tau}-\frac{dn^1(\tau)}{d\tau}\right)+2\alpha
,\label{e20}\\
\frac{dJ^{13}(\tau)}{d\tau}&=&m\frac{dn^3(\tau)}{d\tau}-j^3(\tau),\label{e21}\\
\frac{dJ^{14}(\tau)}{d\tau}&=&m\frac{dn^4(\tau)}{d\tau}-j^4(\tau),\label{e22}\\
\frac{dJ^{23}(\tau)}{d\tau}&=&2\sigma g(u)n^3(\tau)-2\sigma f(u)
n^4(\tau)+m\frac{dn^3(\tau)}{d\tau}-j^3(\tau)
,\label{e23}\\
\frac{dJ^{24}(\tau)}{d\tau}&=&-2\sigma g(u)n^4(\tau)-2\sigma
f(u)n^3(\tau)+m\frac{dn^4(\tau)}{d\tau}-j^4(\tau)
,\label{e24}\\
\frac{dJ^{34}(\tau)}{d\tau}&=&0.\label{e25}
\end{eqnarray}
Now, comparing (\ref{e16}), (\ref{e17}), and (\ref{e27}) we get
$\beta=-\alpha$. If we use the same parameter to describe both of
the world-lines, then $n^t=0$. So we set $n^1(\tau)=-n^2(\tau)$
which is consistent with (\ref{e41}). Taking this into account,
equations (\ref{e13}) and (\ref{e20}) give us
$-n^1(\tau)=n^2(\tau)=\frac{\alpha}{m}\tau+\gamma$ in which
$\gamma$ and $\alpha$ are the initial values of $n^2$ and
$\frac{dn^2}{d\tau}$ respectively. The transverse components of
$n^\mu$ can be obtained from equations (\ref{e68})-(\ref{e69}),
(\ref{e18}), (\ref{e19}), and(\ref{e21})-(\ref{e25}) if one knows
the second particle's world-line and spin orientation. Two
interesting cases are as follow.

Case 1: The second particle moves on a nearby geodesic. We set
$J^{13}(\tau)=0, J^{14}(\tau)=0$. Thus
$j^3(\tau)=m\frac{dn^3(\tau)}{d\tau},
j^4(\tau)=m\frac{dn^4(\tau)}{d\tau}$ and
$J^{23}(\tau)=-2\sigma\frac{dn^4(\tau)}{d\tau},
J^{24}(\tau)=2\sigma\frac{dn^3(\tau)}{d\tau}.$ It follows that
\begin{eqnarray*}
\frac{d^2n^3(\tau)}{d\tau^2}&=-f(u)n^3(\tau)-g(u)n^4(\tau),\\
\frac{d^2n^4(\tau)}{d\tau^2}&=f(u)n^4(\tau)-g(u)n^3(\tau),
\end{eqnarray*}
as we expected.

Case 2: The second particle's spin is such that
$j^3(\tau)=j^4(\tau)=0,$ and $J^{13}(\tau)=mn^3(\tau),
J^{14}(\tau)=mn^4(\tau).$ It follows that
\begin{eqnarray}
\frac{dn^3(\tau)}{d\tau}&=&\frac{2\sigma}{m}\left(f(u)n^4(\tau)-g(u)n^3(\tau)\right),\label{e28}\\
\frac{dn^4(\tau)}{d\tau}&=&\frac{2\sigma}{m}\left(f(u)n^3(\tau)+g(u)n^4(\tau)\right),\label{e29}
\end{eqnarray}
which can be solved for $n^3(\tau), n^4(\tau)$ if $f(u),g(u)$ is
given explicitly in terms of $u=\tau$.
\section{Conclusions}
The equations we have found describe the world-line deviations in
the framework of MPD equations determining the effect of the
spin-curvature coupling on relative accelerations of nearby
particles. They may be helpful in different applications including
the study of chaotic behaviour of spinning particles in certain
space-times \cite{ma}. Another important application of these
equations is to find approximate solutions to MPD equations by
using a known solution, an example is the case two of the previous
section. We applied the equations to the case of motion in the
space-time of a gravitational wave and showed that they can be
integrated analytically. A linear approximation of these equations
may be useful in some applications. In more complex situations
they may be solved at least numerically. In situations in which a
higher accuracy is needed one can use more sophisticated equations
containing higher orders of $\lambda$ in a systematic way. If the
particles experience extra forces like the Lorentz force, the
equations can be modified by adding suitable terms. The example of
the previous section deserves a more extended study.

\vspace{5mm}

\noindent{\bf Acknowledgement}

This work was completed with the financial support of Payame Noor
University research council .

\end{document}